# Background2Background: a self-supervised background removal framework for high dimensional microscopy.

Guangrui Ding[1,5,#], Haonan Lin[2,#], Rong Tang[1,5], Bethany Weinberg[3,5], Jianpeng Ao[1,5], Qing Xia[1,5], Guorong Hu[1,5], Dingcheng Sun[3,5], Jiaze Yin[1,5], Lei Tian[1,3,5], Ji-Xin Cheng[1,3,4,5*]

[1] Department of Electrical and Computer Engineering, Boston University, Boston, MA, USA, 02215

[2] The Wallace H. Coulter Department of Biomedical Engineering, Georgia Institute of Technology and Emory University, Atlanta, GA, 30332

[3] Department of Biomedical Engineering, Boston University, Boston, MA, USA, 02215

[4] Department of Chemistry, Boston University, Boston, MA, USA, 02215

[5] Photonics Center, Boston University, Boston, MA, USA, 02215

[#] Contributing equally

[*]Corresponding author: jxcheng@bu.edu

## Abstract

High-dimensional microscopy captures features across space, time, and spectrum, but complex backgrounds obscure signals. We present Background2Background (B2B), a self-supervised background removal framework leveraging the background's larger correlation relative to the signal. Local pixel shuffling disrupts signal correlations while preserving background structure, transforming signals into effective "noise" for Noise2Noise-based background removal. Across fluorescence, stimulated Raman scattering and mid-infrared photothermal imaging, B2B achieves over 40-fold background suppression and reveals hidden signatures.

## Main

Modern optical microscopy captures biological information across multiple dimensions, including depth, time, and spectrum. Volumetric and temporal fluorescence imaging[1] reveal three-dimensional cellular architecture and dynamic processes, while hyperspectral vibrational imaging[2] maps biochemical heterogeneity through molecular spectral signatures. However, these multi-dimensional datasets often contain complex backgrounds that vary across space, time, and spectrum. Such backgrounds can overwhelm signals and distort quantitative measurements, limiting the biological insights that can be reliably extracted.

Computational background removal can improve image contrast without increasing system complexity or acquisition time. Model-based methods[3-6] generally assume that background is confined to low spatial frequencies and require explicit modeling of the background frequency distribution. Such modeling becomes increasingly challenging in high-dimensional microscopy where background is heterogeneous in different domains. Learning-based algorithms offer greater flexibility but require paired training data and simulated backgrounds for supervised learning[7,8]. Mismatches between simulated and experimentally measured data can therefore limit their performance and generalizability.

Here, we introduce Background2Background (B2B), a self-supervised framework for removing heterogeneous backgrounds in high-dimensional microscopy. B2B exploits differences in correlation length between biological signals and backgrounds to recast background estimation as a denoising problem. It generates a pair of locally randomized images from the input, disrupting the short-range correlations of signals while preserving the longer-range correlations of the background. The pair thus shares background structure, with randomized signals acting as decorrelated "noise." A network trained to predict one image from the other using Noise2Noise (N2N)[9] learns to recover the sSared background without requiring ground-truth. The estimated background is then used to correct the original images.

B2B enables label-free stimulated Raman scattering (SRS)[10] mapping of carbohydrates and nucleic acids in the fingerprint window, high-throughput mid-Infrared photothermal (MIP)[11] imaging of protein and lipid distribution in *C. elegans*, tracking of actin dynamics in wide-field fluorescence microscopy, and one-photon calcium imaging of neuron activity in a mouse brain.

We first explored the physical basis for distinguishing background from biological signals. In wide-field fluorescence, defocus broadens the point spread function (PSF), causing fluorescence from out-of-focus structures to spread over a larger lateral area. Together with diffusely distributed autofluorescence, this produces a spatially varying background that obscures in-focus signals (**Fig. 1a, b**). SRS is accompanied by competing optical heterodyne processes such as cross-phase modulation (XPM) with broad spectral response, while MIP contains broad water absorption peaks (**Fig. 1c**). These background sources are diffusely distributed throughout the sample, giving rise to spatially broad background signals. In general, the background exhibits a longer correlation length than signals (**Fig. 1d**), which builds a foundation for background elimination.

To implement B2B, we introduce local randomization (LRAND), which replaces each pixel with a randomly selected pixel within a local cube (**Fig. 1e**). Enlarging the cube size perturbs structures at progressively longer correlation scales, thereby increasing the Fourier-domain difference between the original and randomized stacks. Changes in the curvature of this difference therefore reflect transitions between characteristic scales. We use the second zero crossing of its second derivative as an estimate of the transition to the longer-range background and select a cube size that disrupts local signal correlations while largely preserving the background (**Fig. 1f** and **Fig. S1**). We then use a pair of independently randomized images to train a U-Net[12] within the N2N framework to estimate the background (**Fig. 1g**). Finally, we extract the signal by subtracting the estimated background from the raw images or dividing the raw images by it (**Fig. 1h**).

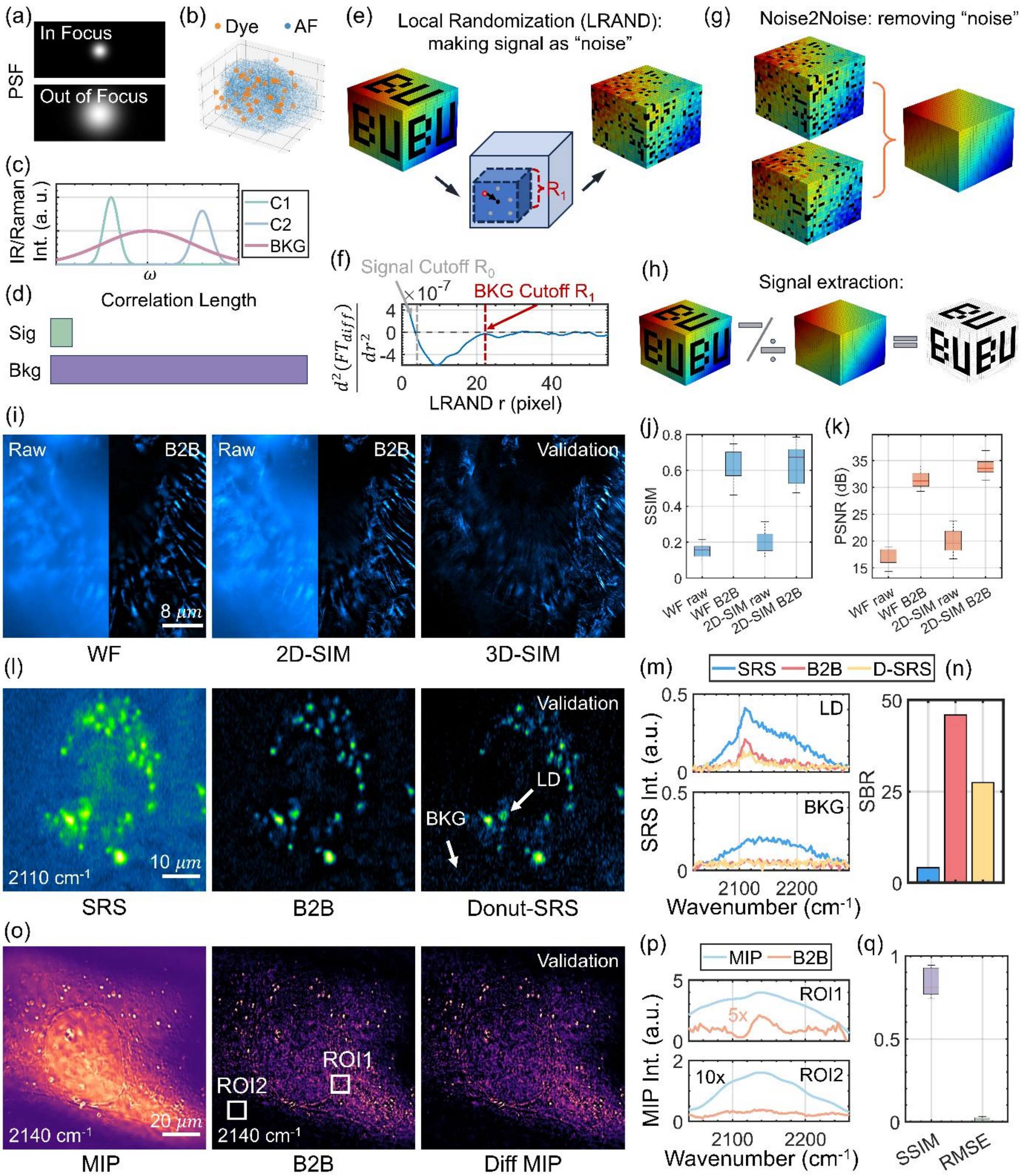

**Figure 1. Principle and validation of B2B.**

(a) Comparison of in-focus and out-of-focus PSFs. (b) Distribution of target dye and autofluorescence (AF). (c) Representative spectral profiles of two chemical bands (C1 and C2) and a broader background (BKG). (d) Schematic comparison of the characteristic correlation lengths of signal and background.

(e) Local randomization (LRAND) operation. (f) Selection rule of optimal cube size. (g) N2N to estimate the background. (h) Signal extraction.

(i) Raw and B2B-processed wide-field (WF) and 2D structured illumination microscopy (2D-SIM) images, with 3D-SIM as the reference. (j, k) SSIM and PSNR evaluation.

(l) Conventional SRS, B2B-processed SRS and donut-SRS reference images. (m) Spectrum of lipid droplet (LD) and background region. (n) SBR evaluation.

(o) Single-wavenumber MIP, B2B-processed MIP, and differential-MIP reference images. (p) Spectrum of selected ROIs. (q) SSIM and RMSE evaluation.

We validated B2B using volumetric fluorescence, hyperspectral SRS, and hyperspectral MIP imaging. For fluorescence, B2B was applied to widefield and 2D structured illumination microscopy (SIM) stacks, achieving 75x and 60x background removal respectively. We used 3D-SIM as the reference[6] (**Fig. 1i**) and quantified improvements using the structure similarity index measure (SSIM; **Fig. 1j**) and peak signal-to-noise ratio (PSNR; **Fig. 1k**). We next evaluated B2B on silent region hyperspectral SRS imaging of PA-d31 treated cell (**Fig. 1l**), using donut-SRS[13] as a reference. B2B reduced the background 40-fold and recovered the PA-d31 spectrum (**Fig. 1m**). It achieved a signal-to-background ratio (SBR) of 46, compared with 27 for donut-SRS (**Fig. 1n**). This advantage reflects B2B's adaptive removal of spatially heterogeneous XPM backgrounds, whereas donut-SRS applies uniform subtraction. B2B also outperformed rolling ball and sliding paraboloid algorithms (**Fig. S2**), which failed to recover signals in lipid rich cytoplasm area (**Fig. S3**). Finally, we evaluated B2B using hyperspectral MIP imaging of azide treated cell, which provides an isolated Infrared (IR) peak in the cell silent window. A previously reported differential MIP approach based on on- and off- resonance measurements served as the reference[14] (**Fig. S4**). B2B enabled background reduction over 150 times and closely reproduced the reference chemical maps and spectrum (**Fig. 1o, p**), as further statistically quantified by SSIM and root mean square error (RMSE) (**Fig. 1q**).

Because B2B does not require high-SBR training data, it is well suited for discovery-driven research, as demonstrated by label-free fingerprint SRS imaging of carbohydrates in single cells. Raman bands near 1100 $cm^{-1}$ provide distinctive chemical contrast but have low cross sections and are often obscured by noise and background. We therefore combined the self-supervised denoiser SPEND[15] with B2B to recover these weak features. The 1048 $cm^{-1}$ band, assigned to C-O stretching in carbohydrates, primarily from glycogen[16] (**Fig. 2a, b**), revealed distinct spatial distributions from that at the C-C skeletal stretching band near 1125 $cm^{-1}$ (**Fig. 2c**). Carbohydrate-rich pools were located near lipid droplets, consistent with electron microscopy observations[17]. Spectra from these regions are shown in **Fig. 2d**. SPEND alone reduced random noise but did not resolve these spatial and spectral features (**Fig. S5** and **S6**). The peak near 1090 $cm^{-1}$ in the nucleolus (**Fig. S7**) is attributed to $-PO_2$[18] associated with nucleic acid and other phosphate-containing biomolecules. The SRS images in the high-wavenumber C-H window provided a morphology reference (**Fig. S8**).

Hyperspectral MIP mesoscope provides high-throughput biochemical imaging over large field of view (FoV)[19]. However, nonuniform illumination and water absorption introduce spatially and spectrally varying backgrounds that obscure chemical contrast. B2B removed this background across an entire worm, revealing protein (amide II and amide I) and lipid (ester band) distributions (**Fig. 2e**) and their overlay (**Fig. 2f**), with clearly resolvable spectral bands (**Fig. 2g, h**). B2B thus combines high-throughput mesoscopic imaging with local molecular specificity to produce chemically interpretable maps of large samples.

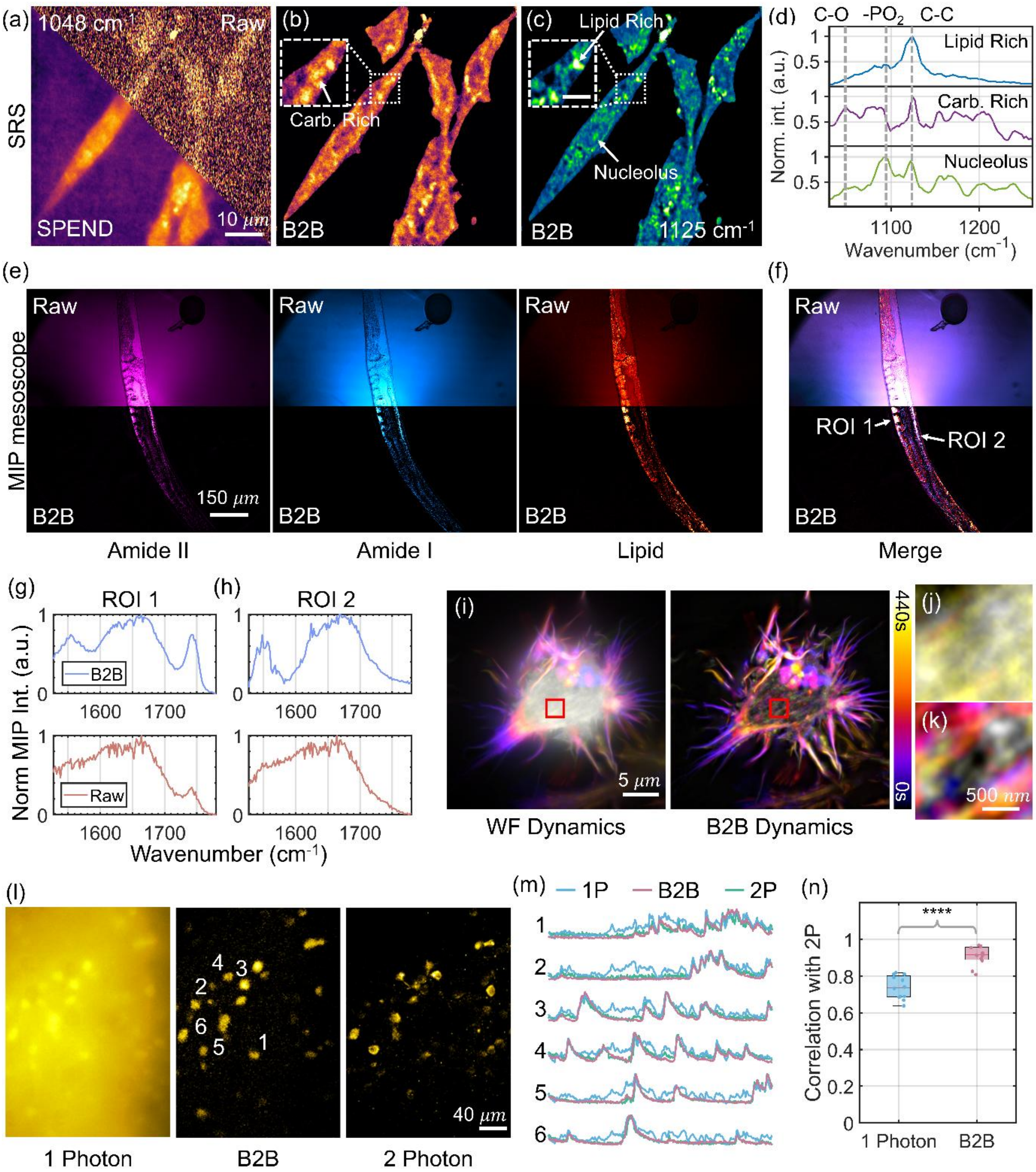

**Figure 2. Applications of B2B to SRS, MIP, and fluorescence modalities.**

(a) Raw and SPEND denoised SRS image at 1048 cm$^{-1}$. (b) B2B processed SRS image at 1048 cm$^{-1}$. (c) B2B processed SRS image at 1125 cm$^{-1}$. Magnified views of the boxed region at 1048 cm$^{-1}$ and 1125 cm$^{-1}$ are shown. Scale Bar: 3 $\mu m$. (d) B2B-corrected spectra from lipid-rich, carbohydrate (Carb.) -rich and nucleolus region.

(e) Raw and B2B-processed Amide II, I and lipid channels. (f) Merged chemical map after B2B processing. (g, h) Spectra from ROI 1 and 2 before and after B2B processing.

(i) Temporal color-coded projections before and after B2B processing. (j, k) Magnified views of the indicated region before and after B2B processing.

(l) Max intensity projection of single photon, B2B processed and two photon calcium imaging. (m)

Temporal profiles of selected neurons. (n) Pearson correlation analysis of single photon and B2B results. P=4.9e-7. 1P: 1 photon. 2P: 2 photon.

Wide-field fluorescence microscopy is widely used to study biological systems, but out-of-focus emission and autofluorescence can obscure signals. Here, we applied B2B to wide-field recordings of propagating actin waves (**Fig. 2i**), revealing their spatiotemporal dynamics[5]. A representative frame is shown in **Fig. S9**. Temporal features buried by background were visible after B2B (**Fig. 2j, k**). Finally, we applied B2B to retrieve neuronal activities in wide-field calcium imaging[7]. B2B improved the localization of active neurons (**Fig. 2l**) and recovered their temporal activity profiles (**Fig. 2m**). Correlation analysis showed increased agreement with the two-photon reference, further supporting B2B's ability to recover neuronal activity (**Fig. 2n**).

Together, these results establish B2B as a broadly applicable framework for background suppression across diverse high-dimensional images, enabling applications ranging from label-free chemical mapping to fluorescence imaging of biological dynamics. The current implementation uses a deterministic estimator and does not explicitly model uncertainty in the recovered background. Incorporating probabilistic or generative models may further improve robustness in datasets with more complex or overlapping signal and background distributions.

## Methods

### Hyperspectral SRS microscope

A custom-built hyperspectral stimulated Raman scattering microscope is used to perform hyperspectral SRS imaging. A femtosecond pulse laser source (Insight, DeepSee+, spectra-Physics) operating at 80MHz provided two synchronized outputs: a tunable pump beam ranging from 680nm to 1300nm and a fixed stokes beam at 1040nm. The pump beam is tuned to 800nm for the C-H region, to 931nm for 1100 $cm^{-1}$ fingerprint region. The Stokes beam is modulated at 2.5MHz by an acoustic optical modulator (1205-C, Isomet) and chirped with a 15cm glass rod (SF57, Schott) before the merging of two beams. After spatial and temporal recombination, the two beams were further chirped using five glass rods to obtain picosecond pulse durations. A motorized delay stage controlled the time delay between the pump and Stokes pulses, thereby tuning the Raman shift through spectral focusing.

The collinear beams were scanned using a 2D Galvo scanner (GVS102, Thorlabs) and focused onto the sample through a 60X water immersion objective (NA=1.2, UPlanApo/IR, Olympus). The transmitted light was collected by an oil condenser (NA=1.4, U-AAC, Olympus). A photodiode (S3994-01, Hamamatsu) is used to collect signals after blocking the Stokes beam. Stimulated Raman gain signals were demodulated at 2.5 MHz using a lock-in amplifier (UHFLI, Zurich Instrument).

### MIP Mesoscope

Large-Field of View (FoV) MIP imaging was performed using a previously reported lab-built MIP mesoscope[19]. The MIP mesoscope operates in a co-propagating configuration, in which the mid-infrared pump and visible probe beams are combined and focused onto the sample through the same objective. Mid-IR excitation is provided by a QCL (MIRcat 2400, Daylight Solutions) outputting pulses at 600 kHz with 300 ns pulse duration, which induces a transient thermal modulation. The IR beam is combined with a 532-nm continuous-wave visible probe beam (Samba 532, Cobolt) via a dichroic mirror. The combined beam is scanned by an X–Y pair of galvanometric mirrors (Saturn 1B, ScannerMAX). A pair of concave mirrors ($f_{SL} = 200\,mm$, CM508-200-P01, Thorlabs; $f_{TL} = 500\,mm$, CM750-500-P01, Thorlabs) relay the scanned beam to the back pupil of a 15×/0.4 NA reflective objective (50105-01, Newport), expanding the beam diameter by 2.5× to fill the objective aperture. The all-reflective scanning relay eliminates chromatic aberrations, enabling IR spectroscopic imaging without re-alignment across visible and IR wavelengths. The probe beam, carrying the photothermal modulation, is collected by a 20×/0.45 NA refractive objective (AMEP4763EO, Olympus) with an axial offset and detected by an amplified photodiode (PDA100A2, Thorlabs). A bias tee (ZFBT-4R2GW+, Mini-Circuits) separates the DC and AC components of the photodiode output. The AC signal is demodulated using a lock-in amplifier (HF2LI, Zurich Instruments) to recover the photothermal signal.

### B2B training and prediction

To enhance the robustness of our model, we augment the datasets by 4-fold through flipping and rotating at 180 degrees. After augmentation, the training set is comprised of 36 stacks, with 10% for validation and 90% for training. A 4-layer Unet architecture based on the CSBDeep framework is employed. The training was conducted on a commercial graphics processing unit (GPU, RTX 4090, Nvidia). Training time is different based on different datasets.

### Open-source data

WF fluorescence, 2D-SIM, 3D-SIM data in **Fig. 1i** is from Dark-based Optical sectioning[6]. SRS dataset in **Fig. 1l** is from Donut-Shaped detection SRS[13]. MIP imaging of Azide in **Fig.1o** is from MIP imaging of Azido photothermal probe (App)[14]. WF fluorescence video in **Fig. 2i** is from Background Filter SIM[5]. Calcium imaging of neuron in **Fig. 2l** is from DeepWonder[7].

### Sample preparation

U87 cells used in **Fig. 2a-d** were seeding in a 35mm glass-bottom dish overnight. Cells were incubated within the medium that contains analogs at a concentration of 20$\mu M$ for 48 hours. Cells were fixed with 10% neutral buffered formalin for 30 minutes and then washed with phosphate-buffered saline (PBS) three times.

C. elegans used in **Fig.2e-h** were cultured on nematode growth medium (for 1L: phosphate buffer 20mM, $CaCl_2$ 0.8mM, $MgSO_4$ 0.8mM, agar 17.5g, NaCl 3.0g, peptone 2.5g, cholesterol 0.005g) and grown at 15°C. Plates were seeded with OP-50 Escherichia coli. For imaging, young adult worms (1-3 days old) were mounted via picking into 1% $NaN_3$ in M9 buffer on a 24x60mm #1 thickness cover slide. Double-sided tape was used as a spacer, and another #1 thickness cover slide was lowered over the top of the tape to sandwich the sample.

### Acknowledgements.

This work is supported by NIH grants R35GM136223, R01 EB032391, and R01 EB035429 to JXC.

**Author contributions.**

G.D., H.L., and J.C. co-designed the project. G.D. proposed the B2B framework. G.D., H.L., R.T., B.W., J.A., Q.X., J.Y., and D.S. conducted the SRS and MIP experiments. G.D. and G.H. conducted the fluorescence imaging experiment. L.T., H.L., and J.C. supervised the project. G.D., H.L., and J.C. wrote the manuscript. All authors read the manuscript.

**Data availability.**

The datasets have been made publicly available on https://github.com/buchenglab/B2B

**Code availability.**

The original codes have been made publicly available on https://github.com/buchenglab/B2B

# Supplementary Information

## Background2Background: a self-supervised background removal framework for high dimensional microscopy

Guangrui Ding[1,5,#], Haonan Lin[2,#], Rong Tang[1,5], Bethany Weinberg[3,5], Jianpeng Ao[1,5], Qing Xia[1,5], Guorong Hu[1,5], Dingcheng Sun[3,5], Jiaze Yin[1,5], Lei Tian[1,3,5], Ji-Xin Cheng[1,3,4,5*]

[1] Department of Electrical and Computer Engineering, Boston University, Boston, MA, USA, 02215

[2] The Wallace H. Coulter Department of Biomedical Engineering, Georgia Institute of Technology and Emory University, Atlanta, GA, 30332

[3] Department of Biomedical Engineering, Boston University, Boston, MA, USA, 02215

[4] Department of Chemistry, Boston University, Boston, MA, USA, 02215

[5] Photonics Center, Boston University, Boston, MA, USA, 02215

[#] Contributing equally

[*]Corresponding author: jxcheng@bu.edu

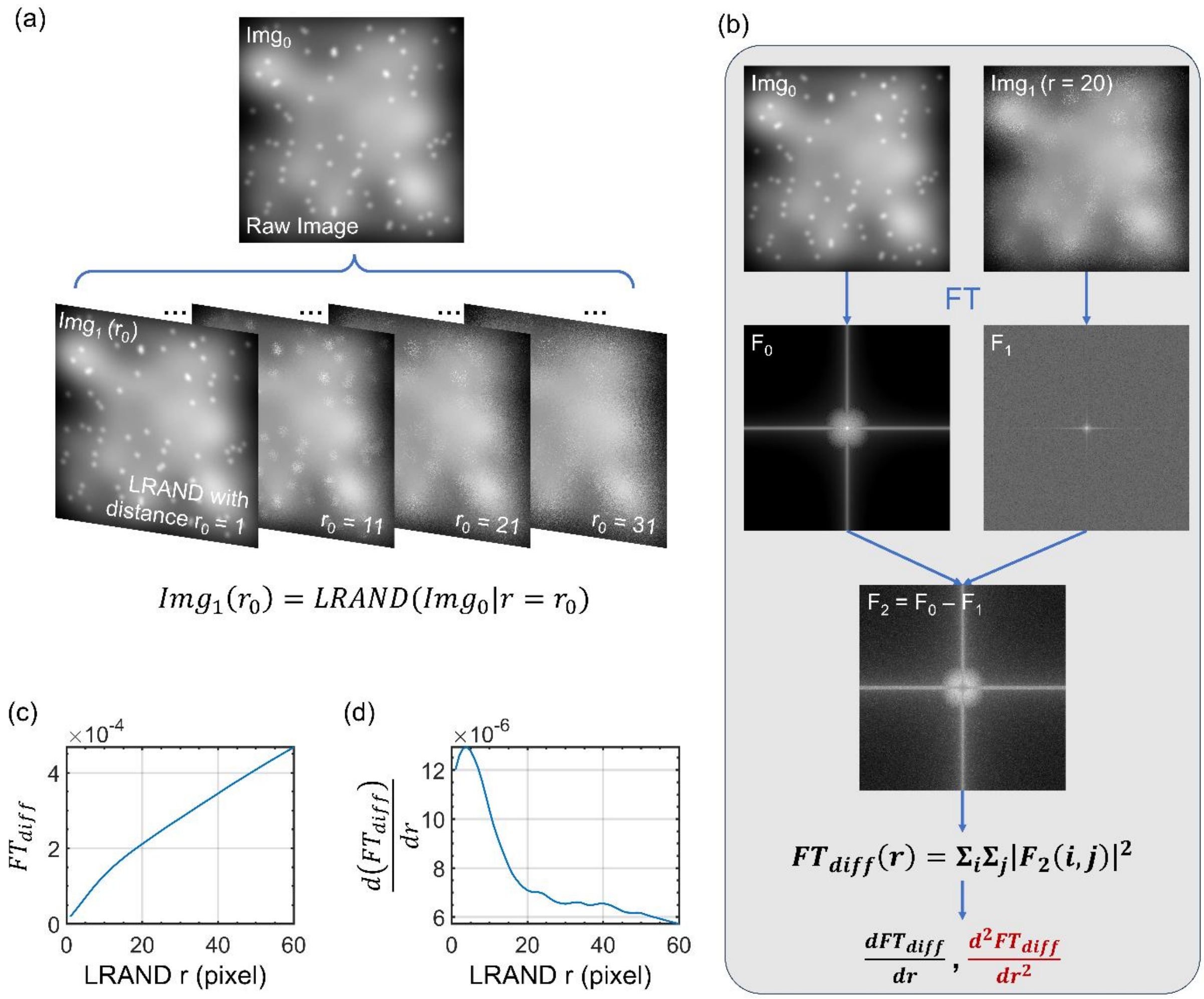

**Figure S1**. **Determination of the optimal LRAND radius $R_1$.** (a) LRAND applied to the raw image using different radius r. (b) Calculation of Fourier domain difference between the original and randomized images, together with its first and second derivative with respect to r. (c, d) Example of Fourier domain difference and its corresponding first derivative.

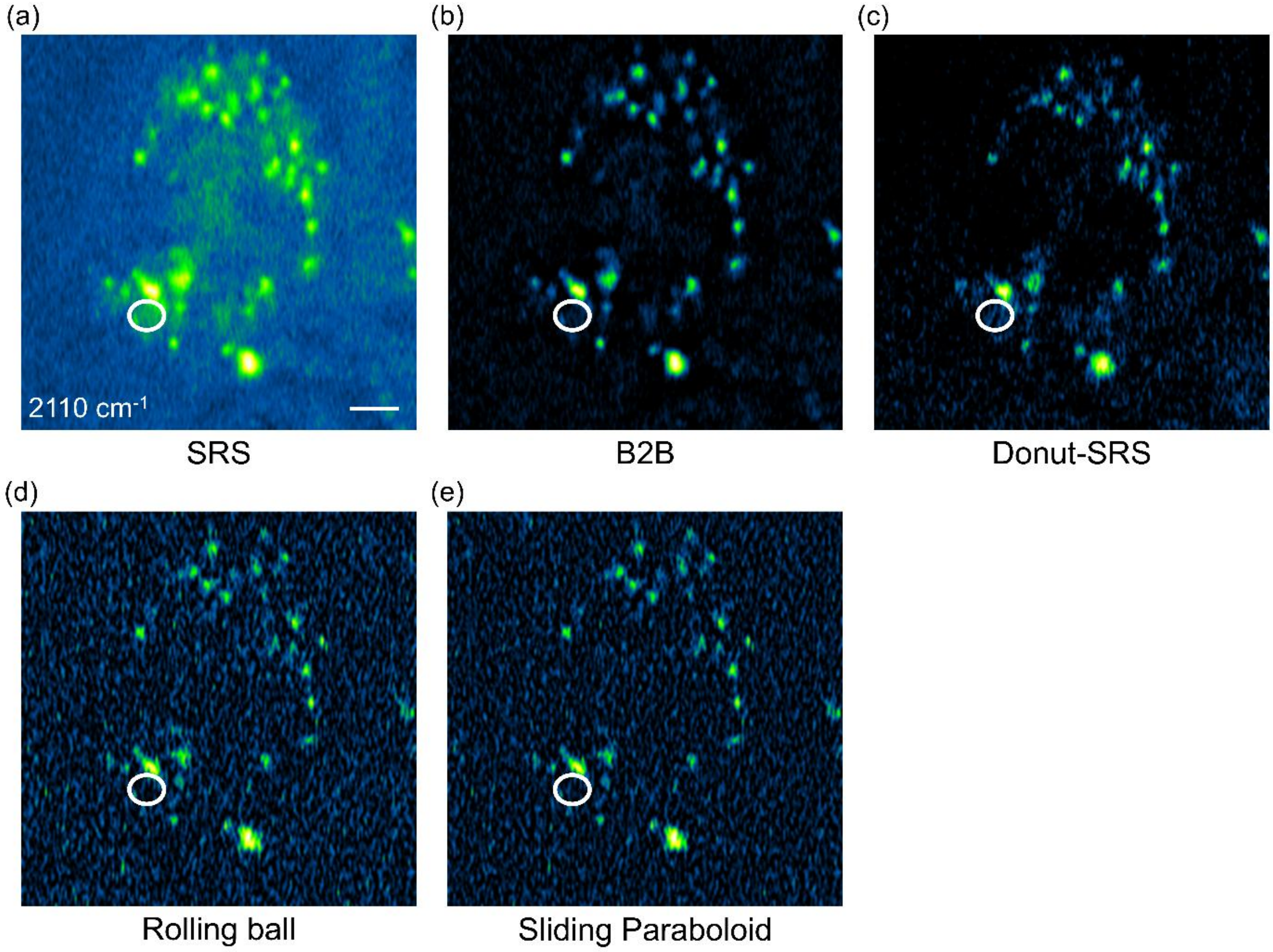

**Figure S2**. **Spatial performance of rolling ball and sliding paraboloid methods.** Scale bar: 10 $\mu m$.

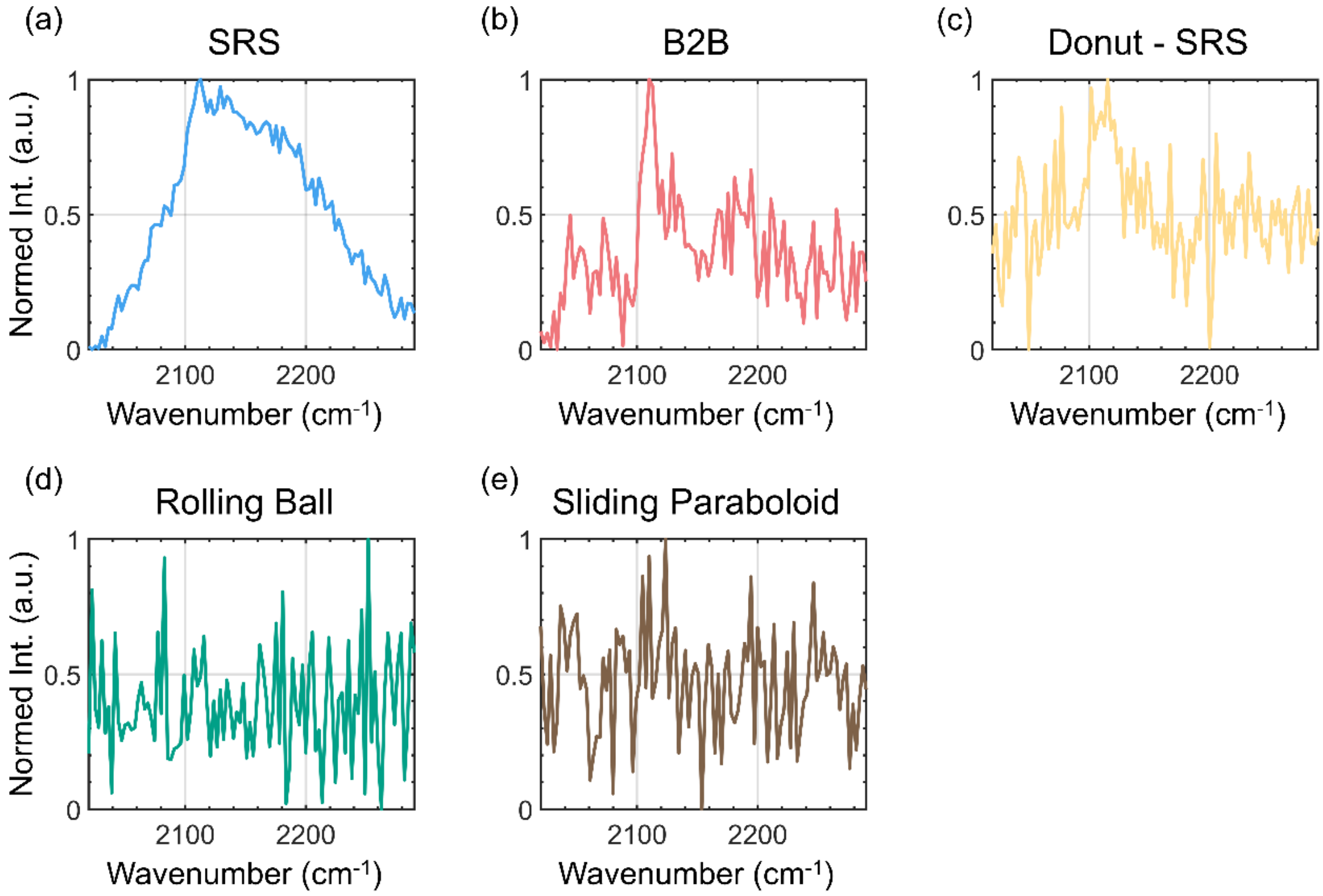

**Figure S3**. **Spectral results of each method.**

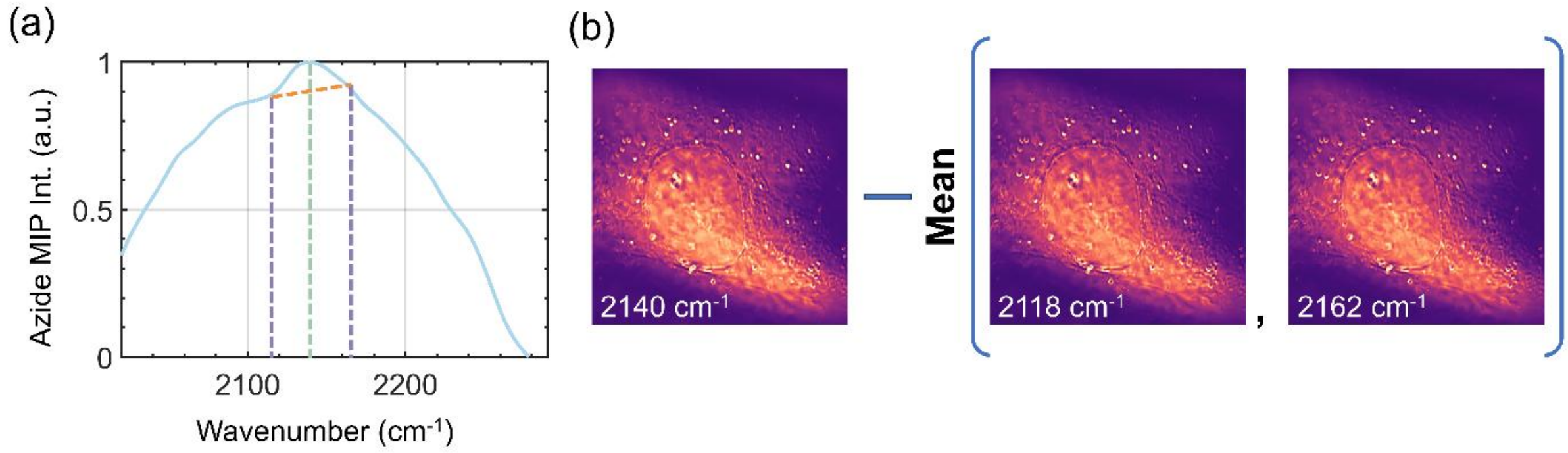

**Figure S4**. **Azide signal extraction via subtraction of images at three different wavenumbers.** (a) Intensity-normalized spectrum of Azide in silent region. (b) subtraction process.

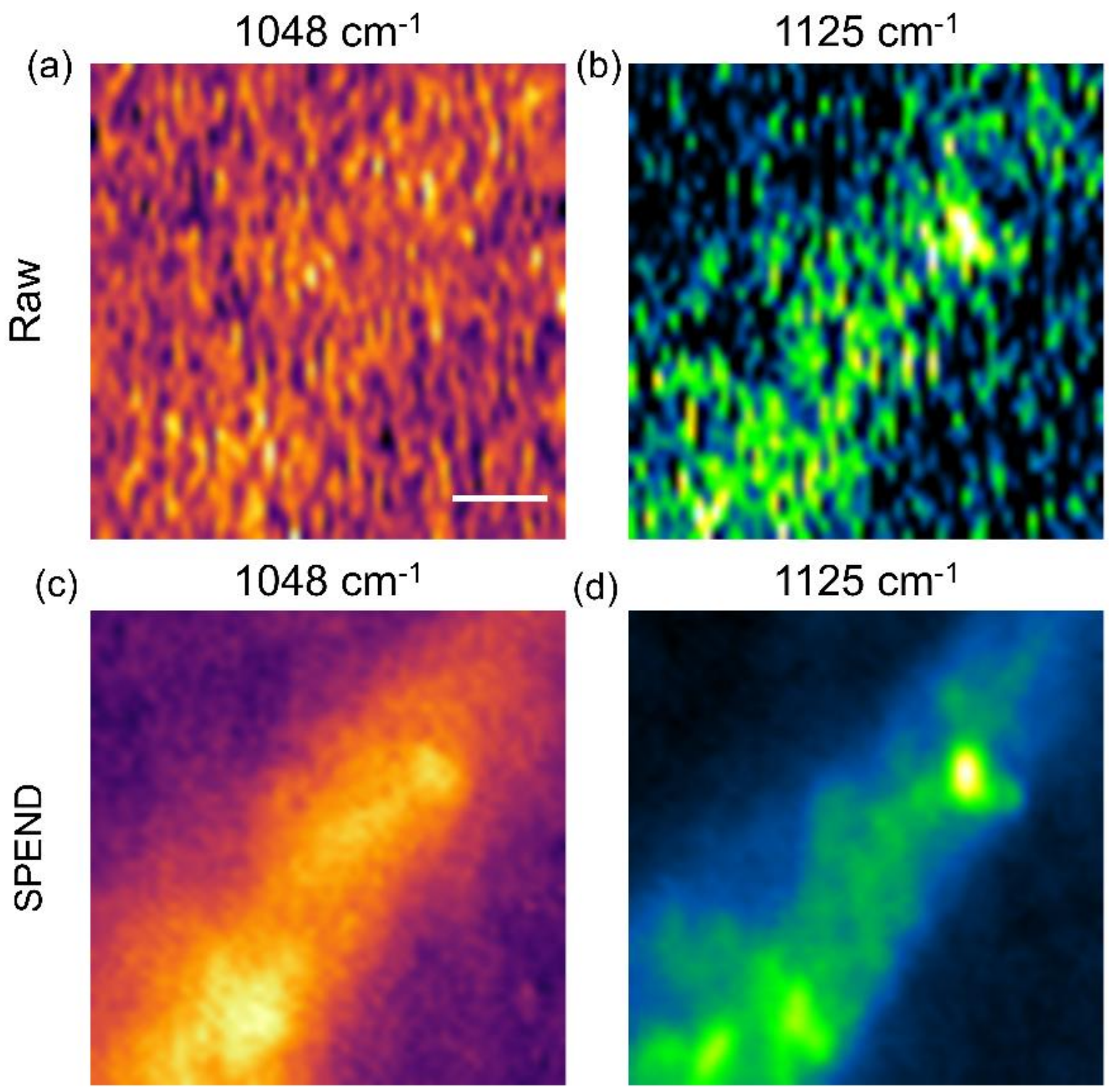

**Figure S5**. **Comparison of raw and SPEND processed SRS image.** (a, b) Raw SRS images at 1048 $cm^{-1}$, (b) 1125 $cm^{-1}$. (c, d) SPEND processed SRS images at (c) 1048 $cm^{-1}$, (d) 1125 $cm^{-1}$. Scale Bar: $1\ \mu m$.

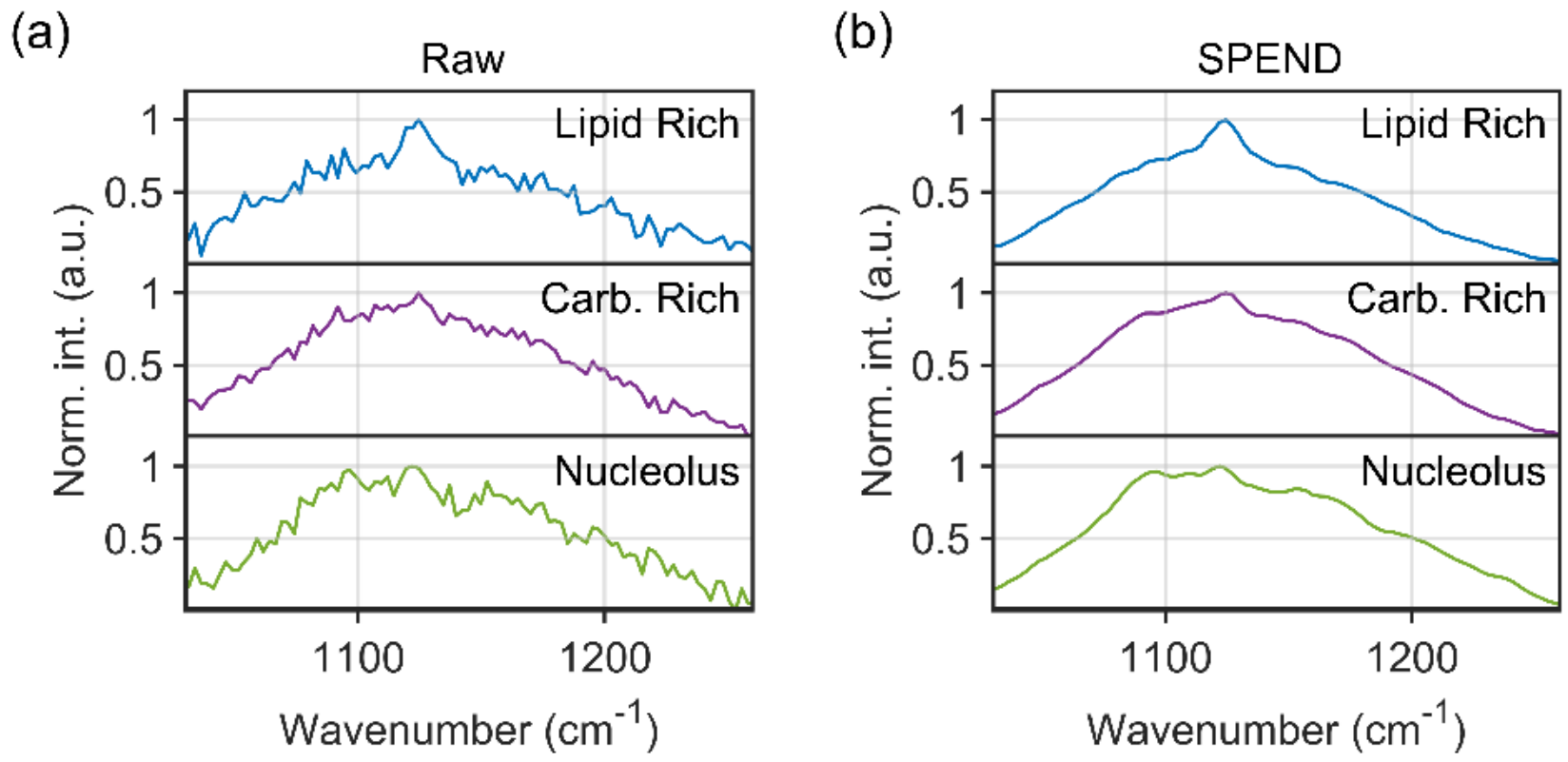

**Figure S6**. **Spectrum profile of selected regions.** (a) Raw. (b) SPEND processed.

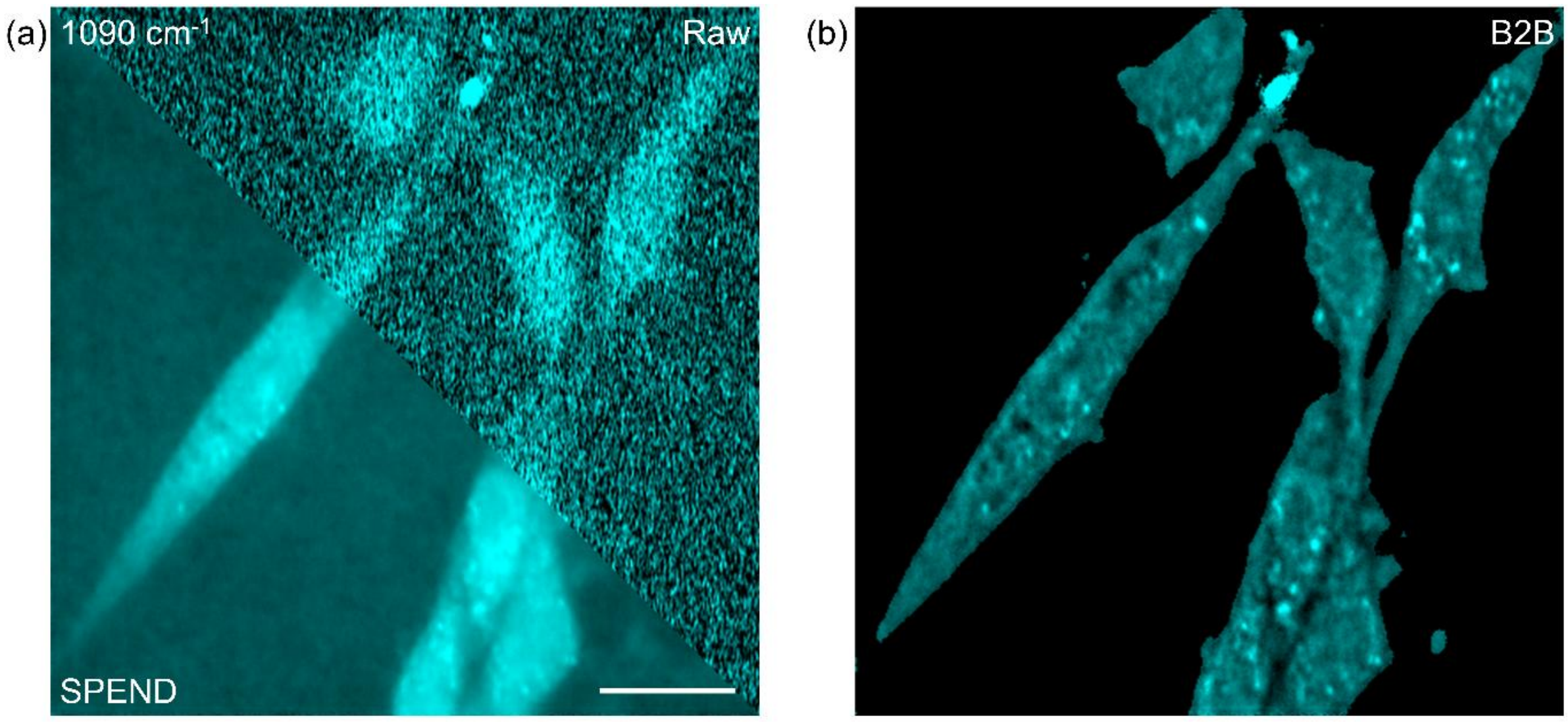

**Figure S7**. **SRS image at 1090 cm$^{-1}$.** (a) Raw and SPEND processed images. (b) B2B processed image. Scale bar: 10 $\mu m$.

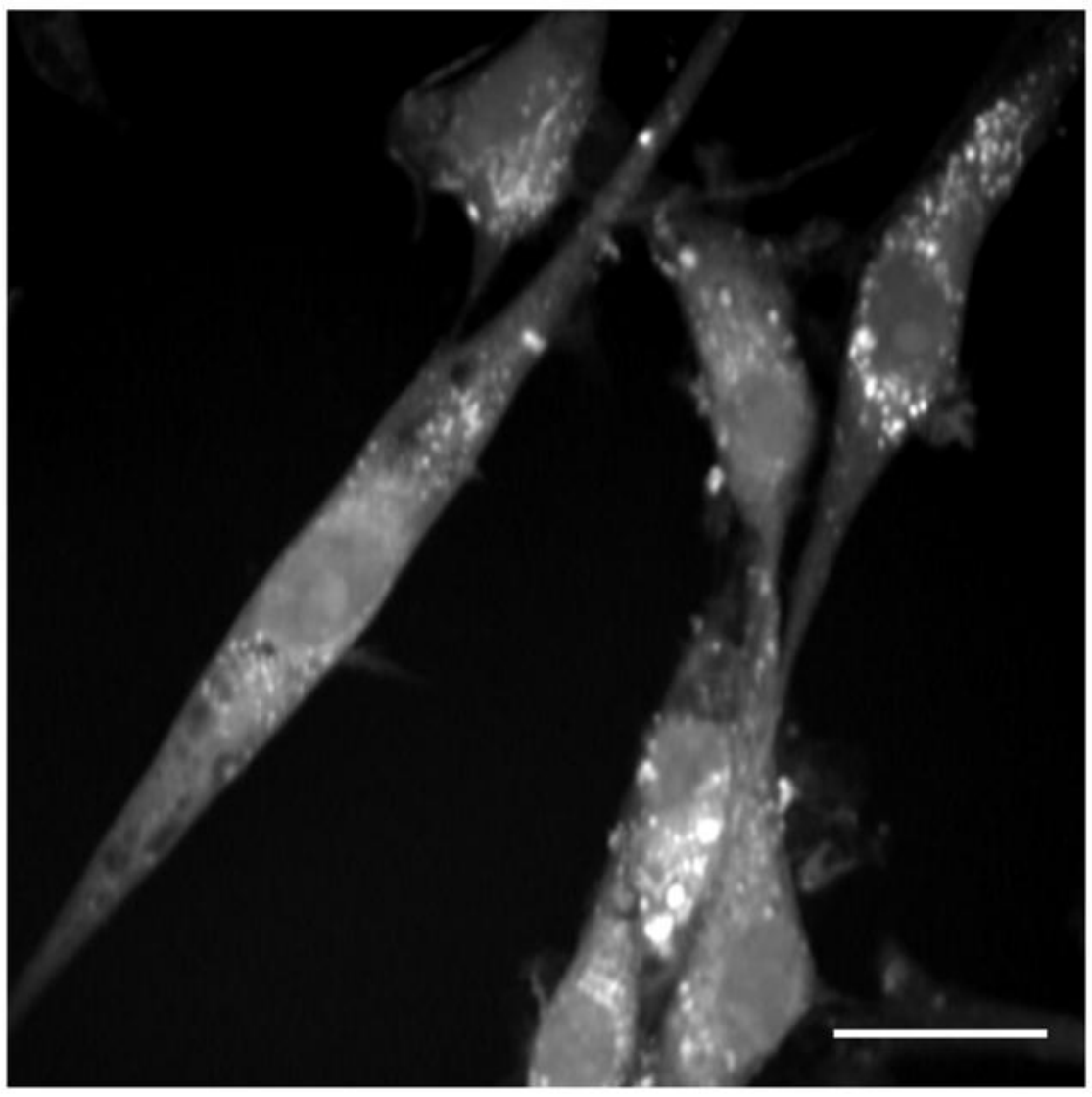

**Figure S8**. **Stack average of hyperspectral SRS imaging of U87 in CH region.** Scale bar: 10 $\mu m$.

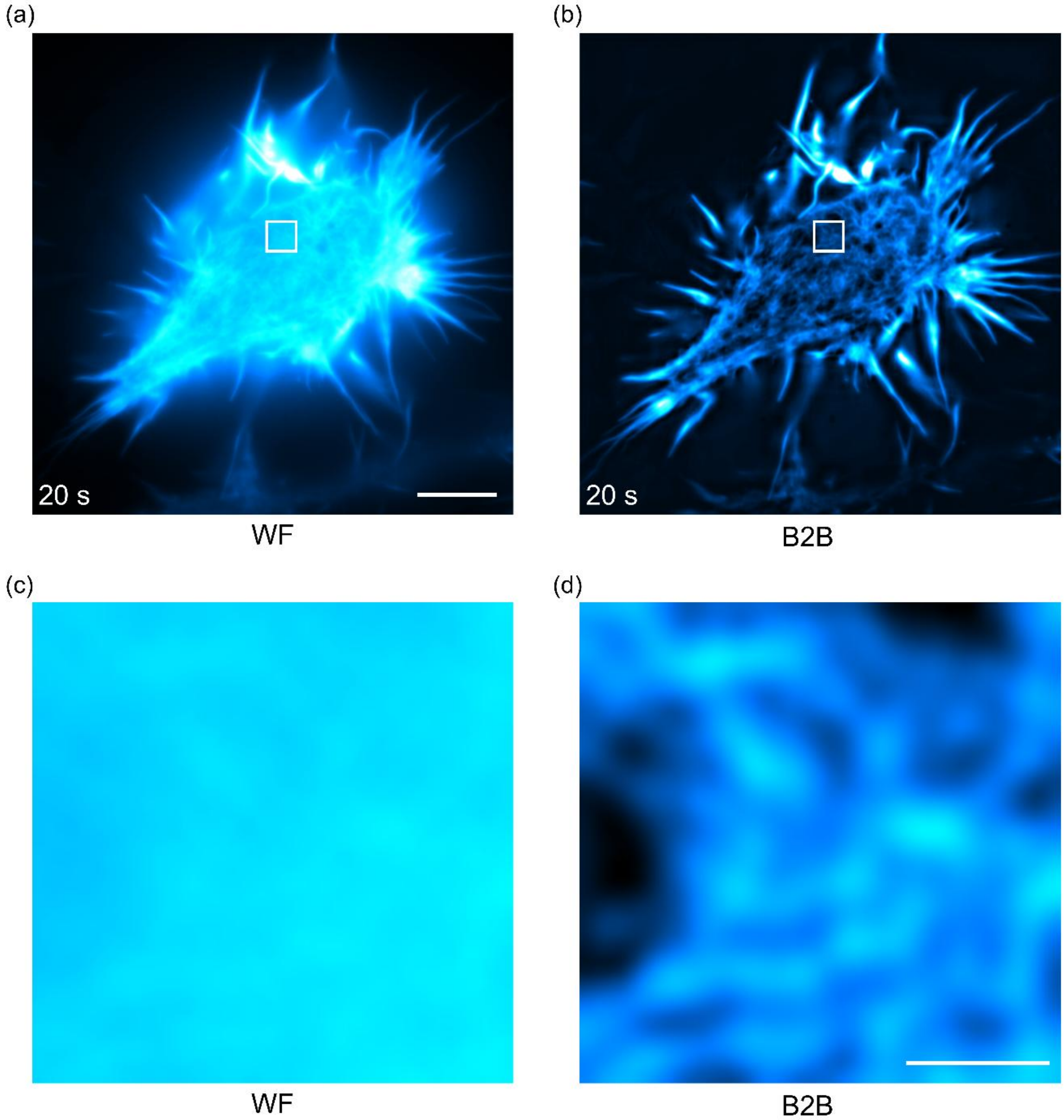

**Figure S9**. **Single frame comparison of widefield fluorescence images before and after B2B processing.** (a) Raw, (b) B2B at 20s. (c, d) Magnified views of selected regions in a and b respectively. Scale bar: (a) 5 $\mu m$. (d) 500 $nm$.